\documentclass[prl,twocolumn,showpacs,letter]{revtex4-1}%

\usepackage{amsmath,amsfonts}
\usepackage{graphicx}
\usepackage{epsfig}
\usepackage{dcolumn}
\usepackage{bm}
\usepackage{times}
\usepackage{epstopdf}
\begin{document}

\title{The Totally Asymmetric Simple Exclusion Process on Networks}
\author{Izaak Neri$^{1}$, Norbert Kern$^{1}$, and Andrea Parmeggiani$^{2}$} 
\affiliation{$^{1}$ Laboratoire Charles Coulomb, UMR 5221 CNRS-UM2, cc 069,
Universit\'e Montpellier II, Pl. E. Bataillon, 34095 Montpellier Cedex
5, France \\ 
${^2}$ Laboratoire de Dynamique des Interactions Membranaires Normales et
Pathologiques, UMR 5235 CNRS-UM2-UM1, cc 107, Universit\'e Montpellier II, Pl. E. Bataillon, 34095 Montpellier Cedex 5, France}
\date{\today}

\begin{abstract}
We study the totally asymmetric simple exclusion process (TASEP) on complex networks, as 
a paradigmatic model for transport subject to excluded volume interactions. 
Building on TASEP phenomenology on a single segment and borrowing ideas from random networks we investigate the effect of connectivity on transport. 
In particular, we argue that the presence of disorder in the 
topology of vertices crucially modifies the transport features of a network: 
irregular networks involve homogeneous segments and have a bimodal distribution of edge densities, whereas regular networks are dominated by shocks leading to a unimodal density distribution. 
The proposed numerical approach of solving for mean-field transport on 
networks provides a general framework for studying TASEP on 
large networks, 
and is expected to generalize to other transport processes.

\end{abstract} 

\maketitle
Delivering matter, energy or information is a crucial requirement for the functioning of any complex system, ranging from the sub-cellular level of biological organisms to globe-spanning  man-made structures. 
Transport is often organized along line-like pathways, which are in turn interconnected to form a network structure. In this perspective, diffusion along networks has been studied extensively (see, e.g., 
\cite{Ves,Lopez}). On the other hand, interactions play a fundamental role in the transport properties of many systems: intracellular traffic of molecular motors on the cytoskeleton, pedestrian traffic on an ensemble of paths and traffic of information packages on the internet are but a few prominent examples. They lead to emergent collective phenomena, such as the appearance of jams. 

The {\it totally asymmetric} {\it simple exclusion process} (TASEP) is a 
paradigmatic model for one-dimensional non-equilibrium transport 
subject to excluded volume interactions: 
entities (``particles'') hop in a given direction, but 
cannot occupy the same place \cite{Schutz}.  
TASEP was initially introduced as a model for the kinetics of RNA
polymerization by ribosomes \cite{Mac68}, but has since then 
received much general interest, including from fundamental
statistical physics \cite{Privman} and mathematics \cite{Ligget}.
Numerous generalizations have been developed and applied to various areas, 
such as the collective motion of motor proteins along cytoskeletal 
filaments, vehicular traffic, etc. \cite{examples,Chow05}.

The collective behavior of exclusive transport {\it on a network}, however, 
is not well understood at this stage. 
A body of numerical work on TASEP-like models on networks, often with 
complex details, exists in the context of traffic \cite{Chow05}, 
but less so on biological transport \cite{Greu10}. 
In terms of a paradigmatic analysis 
based on TASEP, knowledge is still limited to either 
simple topologies
with at most two junctions \cite{junctions, Em09}, 
or involves structureless links, {\it e.g.} tree-like networks.


%

\begin{figure}[ht!]
 \begin{center}
 \includegraphics[width=.40 \textwidth]{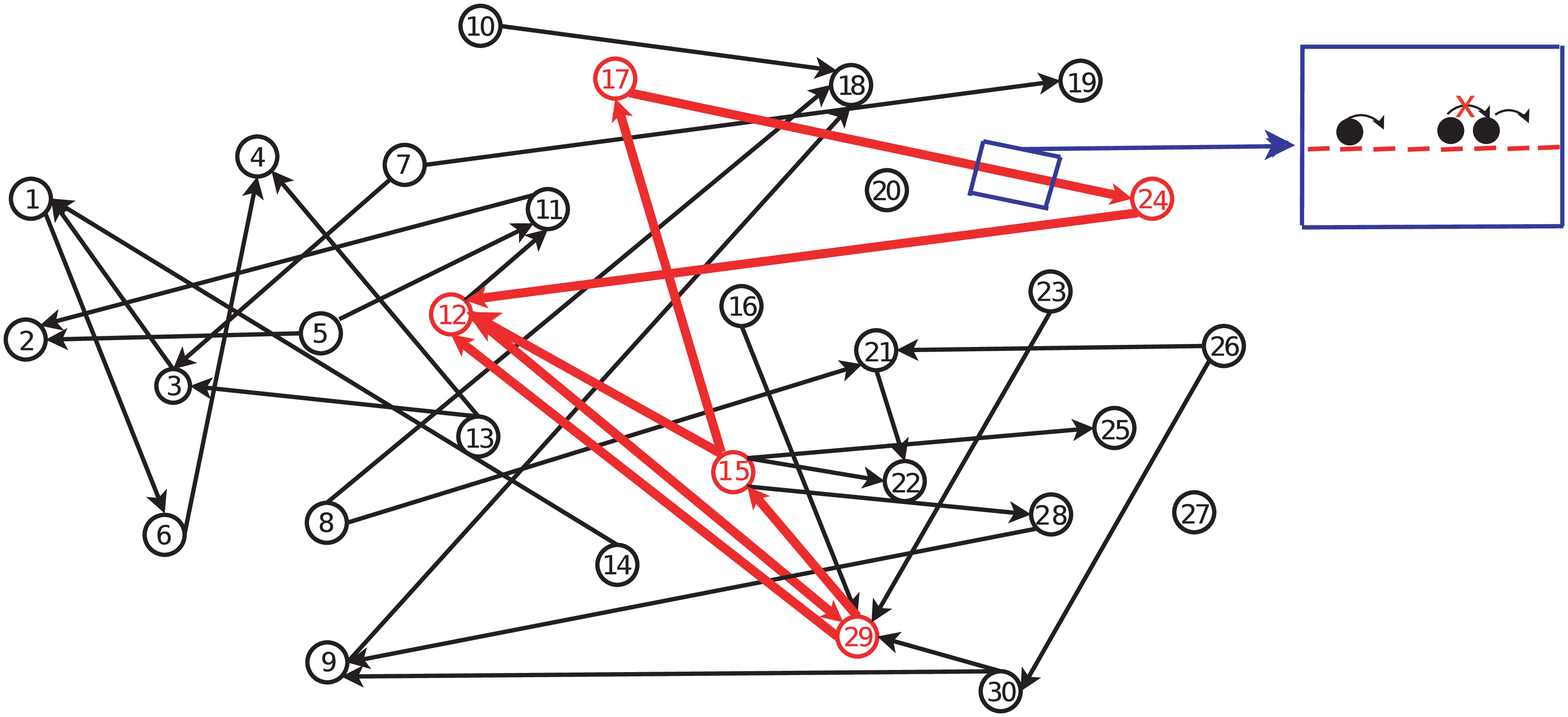}
 \caption{(color online). An example of a Poissonian network of $M=30$ junctions
 with $c=1$ and its strongly connected
 component (bold).  Every segment
 consists of $L$ sites on which particles undergo TASEP
 dynamics.  At the junctions the particles choose one of
 $k^{\rm out}_v$ outlets.}
\label{fig:graph}
 \end{center}
 \end{figure}

In this letter we address the question of how the topology of a network 
affects its TASEP transport characteristics.  
Combining concepts from the area of complex networks \cite{Ves} 
with mean field (MF) methods for TASEP in the presence of 
junctions \cite{Em09} we
construct the global behavior from that of single segments. 
This allows us to rationalize many features of transport on large-scale 
random networks in terms of theoretical arguments, and furthermore 
leads to an algorithm to numerically solve the MF problem of TASEP 
transport on a large scale network. 
In particular we argue that {\it irregularity}, i.e. randomness in the vertex 
degrees, strongly modifies the transport properties of a network. 
\\
\paragraph{TASEP on a network. --}
We generalize the TASEP transport rules \cite{Mac68, Schutz} to a closed network
of $N_{\rm S}$ directed segments and $N_{\rm V}$ vertices or
junctions.  The segments consist of $L$ sites along which particles 
perform unidirectional random sequential hops subject to  
hard-core on-site exclusion. At
the junctions, particles from $k^{\rm in}_v$ incoming segments compete for 
occupying the same vertex site $v$, while a particle can leave the junction through one of $k^{\rm out}_v$ outgoing segments with equal probability.  
Figure \ref{fig:graph} serves as an illustration for a
network with $M\!\!=\!\!30$ junctions. 
We write $\rho_v$ for the average occupancy of a vertex $v$ with $0\!\leq\!\rho_v\!\leq\!1$. 
Particle number conservation in the junction $v$ reads
\begin{eqnarray}
 \frac{\partial \rho_v }{\partial t}
= 
\sum_{v' \rightarrow v} J_{(v',v)}
- 
\sum_{v \leftarrow v'' } J_{(v,v'')} \quad,
\label{eq:massCons}
\end{eqnarray}
the sums run over vertices $v'$ identifying incoming 
segments $(v',v)$ and over vertices $v''$ for 
outgoing segments $(v,v'')$.

We briefly review the behavior of an isolated segment 
linked to reservoirs, which we will build on. 
Its average density $\rho$ and current $J$ 
are known to be homogeneous, provided  that segments are at least of moderate size, such that boundary effects remain small. Both are set by the entry rate $\alpha$ and exit rate $\beta$ \cite{Der92}:
\begin{eqnarray}
  \rho (\alpha,\beta) = 
  \left\{
  \begin{array}{ccccc}
    \alpha  &\ &\alpha\leq \beta, \alpha<1/2  &\ &\mbox{(LD)} \\ 
    1-\beta &\ &\beta\leq \alpha, \beta <1/2  &\ &\mbox{(HD)}\\ 
    1/2     &\ & \alpha,\beta\geq 1/2        &\ &\mbox{(MC)}
  \end{array}\right.
  \label{eq:RMF}
\end{eqnarray}
with the current given by the parabolic current-density relation
\begin{eqnarray}
J(\alpha,\beta)=\rho(\alpha,\beta) \, \big(1-\rho(\alpha,\beta)\big) 
\,.
\label{eq:JMF}
\end{eqnarray}
These three homogeneous phases are {\it high density} (HD), {\it
low density} (LD) and {\it maximal current} (MC), but for
$\alpha=\beta$ a non-homogeneous {\it shock phase} (SP) arises, 
for which LD and HD regions coexist on the same segment, 
separated by a diffusing domain wall
\cite{dwphenomenology}.
\\
\paragraph{Mean-field theory and algorithm. --}
We extend the MF analysis of
\cite{Em09} to large scale networks to establish an algorithm operating on 
the junction occupancies $\rho_v$ as the only variables. 
To do so, we neglect correlations between neighbouring sites, as is usually done in MF \cite{Der92}.
Then the entry (exit) rate $\alpha$ ($\beta$) for a segment
are {\it effective} rates, which depend only on 
the occupancies of the adjacent vertices \cite{Em09}:
\begin{eqnarray}
\alpha = \rho_{v'}/k^{\rm out}_{v'}
\qquad \mbox{and} \qquad 
\beta = 1-\rho_{v''}
\ .
\label{eq:rateMF} 
\end{eqnarray}
The entry rate $\alpha$ is reduced according to the out-degree 
$k^{\rm out}_{v'}$,
since particles on vertex $v'$ are distributed uniformly over all 
outgoing edges.  Assuming homogeneous segments, 
the effective rates in (\ref{eq:RMF}) and
(\ref{eq:JMF}) can be substituted into the
continuity equation (\ref{eq:massCons}) to yield a
closed set of $N_{\rm V}$ equations in the vertex occupancies:
\begin{eqnarray}
\frac{\partial \rho_v}{\partial t} 
= 
\sum_{v' \rightarrow v}   
\! J \!
\left(\!
 \frac{\rho_{v'}}{k^{\rm out}_{v'}},1 \! 
 - 
 \! \rho_v \!\right) 
 - 
 \sum_{v''  \leftarrow v} 
 \!J \!
 \left(\!
 \frac{\rho_{v}}{k^{\rm out}_v},1 \!  - \! \rho_{v''} \!
 \right)
, \ \ \
\label{eq:MF}
\end{eqnarray}
where the sums run over all vertices $v'$ ($v''$) which are 
upstream (downstream) from $v$.
The microscopic dynamics lack particle-hole symmetry at the junctions, as is reflected in Eq. (\ref{eq:MF}) by the factor $1/k^{\rm out}_{v'}$ in the entry rate.  

\begin{figure}
 \begin{center}
 \includegraphics[width=0.40 \textwidth]{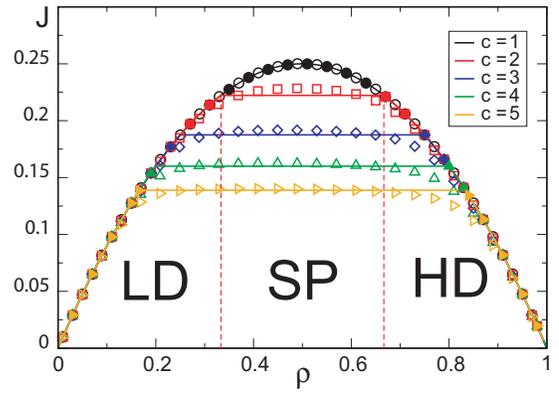}
 \caption{(color online). The average segment current $J$ as a function of the
 overall density $\rho$ on a Bethe network of degree $c$ superposing numerical
 mean-field predictions (closed symbols),  explicit simulations ($N_V=80$
 junctions, $L=100$ sites per edge, open symbols) 
 and a one-vertex analytical result 
 Eq. (\ref{MF:Bethe}) (solid lines).  The dashed lines delimit the 
different phases
 for $c=2$. 
}
\label{fig:JRhoSeg}
 \end{center}
\vspace{-0.5cm}
 \end{figure}

The numerical MF algorithm consists of iteratively finding
the stationary solution to (\ref{eq:MF}), thereby 
achieving considerable computational advantage upon simulations since 
we only need to update the $N_V$ junction occupancies $\rho_v$. 
In the following we will study transport on random networks as model systems, complementing MF solutions of Eqs. (\ref{eq:MF}) by explicit 
simulations.
Throughout, we exploit the fact that macroscopic observables are self-averaging 
on these ensembles, as we have verified directly by analyzing different 
network instances.
\\
\paragraph{TASEP on a Bethe network. --}
As an example of closed random graphs with regular topology we consider the 
{\it directed Bethe network}, drawn from the $c$-regular ensemble \cite{Bol},   
in which all vertices have identical connectivity $c =k^{\rm in}=k^{\rm out}.$
The {\it undirected}  Bethe network is well known in statistical
mechanics from the study of, for example, spin models on graphs \cite{Baxter}.
Figure \ref{fig:JRhoSeg}  shows the average segment current 
$J$ as a function of the overall density $\rho$.
The standard result for the current $J(\rho)\!=\!\rho(1-\rho)$ is 
recovered for $c\!=\!\!1$, where the Bethe network reduces to a simple ring. 
As connectivity is increased, the overall current is {\it reduced}: 
this may appear counter-intuitive, but reflects the fact that 
vertices progressively become bottlenecks and block the flow of particles.
More precisely, the current parabola is truncated 
at intermediate densities by a plateau-like region, which widens but 
lowers with connectivity $c$.
To interpret these results, we first give an explanation based on the 
phenomenology of TASEP on a line. To this end, we point out that
an analytical solution 
to Eqs. (\ref{eq:MF}) can be given for the Bethe network, since all 
$N_V$ equations become identical due to equal vertex connectivities.
Therefore the solution requires identical occupancies 
$\rho_v$ for all vertices, from which segment currents and densities follow:
all vertices, and therefore all segments, are equivalent. 
The transition from the LD to the HD phase
appears when the effective rates (\ref{eq:rateMF}) are equal
($\alpha\!=\!\beta$), i.e. at $\rho_v \! = \! c/(c+1)$.  
Using Eqs. (\ref{eq:RMF}-\ref{eq:rateMF}) 
this leads to distinct regimes for 
the current, yielding the truncated parabola for 
the MF current-density relation:
\begin{eqnarray}
  J\left(\rho\right) 
  =  
  \left\{\begin{array}{ccc}
  \frac{c}{(c+1)^2}    && \mbox{for} \ \  \rho^* < \rho < 1-\rho^*
  \\
  \rho (1-\rho)        && \mbox{otherwise} 
  \end{array}\right. 
  \ \ ,
  \label{MF:Bethe}
\end{eqnarray}
with LD phases (for $\rho<\rho^*$), and HD phases 
(for $\rho>1-\rho^*$), where $\rho^*=1/(c+1)$.
The plateau can be rationalized in terms of domain 
wall phenomenology \cite{dwphenomenology},
recalling that $\alpha=\beta$ indicates a SP with a diffusing domain wall 
between LD and HD zones which coexist on the same segment.
Since both zones have complementary densities ($\rho_{\rm LD} = 1-\rho_{\rm HD}$), 
the current is not affected as further particles are accomodated by growing 
the HD zones at the expense of the LD zones
\cite{dwphenomenology,Em08}. 

These MF arguments capture the essential transport features well, as is 
shown by the simulation data on Fig.\ref{fig:JRhoSeg}
(and this remains true also for triangular and square lattices).
Deviations arise, however, on the current plateau 
(where MF underestimates the current) 
and close to the transitions (where explicit simulations furthermore reveal 
the particle-hole asymmetry, which increases with connectivity $c$).
The numerical MF algorithm does not provide a solution on the plateau, since the  assumption of homogeneous segments does not hold, but it otherwise reproduces the theoretical results with great precision.
In summary, despite its random nature, TASEP transport through a Bethe network may be understood in terms of a single effective vertex, similar to the Ising model on a Bethe lattice \cite{Baxter}. 
\\
\paragraph{TASEP on a Poissonian network. --}

In order to explore the effect of irregularity, i.e. non-uniform vertex 
connectivity, we study TASEP on the Poissonian ensemble, closely related to the
Erd\"os-R\'enyi ensemble \cite{Bol}: any two vertices are connected with 
probability $c/N_V$, yielding an average connectivity $c$.
In order to avoid artefacts we consider transport on the 
{\it strongly connected component} (SCC) of the Poissonian network, 
in which each vertex can be reached from all other vertices.  
We find the SCC using an algorithm developed by Tarjan
\cite{Tar}. For illustration we refer to Fig.\ref{fig:graph}, which shows a
particular realization of a Poissonian ensemble with $c\!=\!1$, the SCC highlighted
in bold. 

\begin{figure}
 \begin{center}
 \includegraphics[width=.40\textwidth]{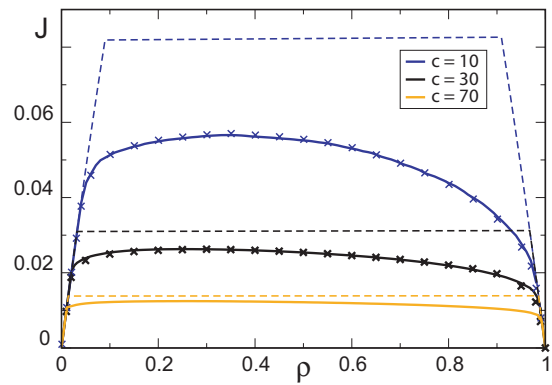}
 \caption{(color online). The current-density characteristic of the strongly connected 
 component of Poissonian graphs of given average connectivity $c$.  
 Simulations (markers)  with $N_{\rm V}=200$ junctions 
 and segments of $L=100$ sites are compared to 
 mean field results (solid lines) for the identical network.
 For comparison, the Bethe result Eq. (\ref{MF:Bethe}) is shown for the 
 same $c$ (dotted lines). 
 Results are {\it not} ensemble-averaged: sample 
 to sample fluctuations are of the order of the symbol sizes, 
but mean field and  simulation results for a given topology coincide as in the example.
 }
 \label{fig:currentDensity}
 \end{center}
\vspace{-0.5cm}
 \end{figure}

We first comment on the transport characteristic $J(\rho)$ of TASEP transport 
on Poisson networks.
Figure \ref{fig:currentDensity} shows the current $J$ 
(averaged over all segments) as a function of 
density $\rho$, for various connectivities $c$. 
The MF results are in excellent agreement with simulations, 
thereby validating the MF algorithm for disordered networks.  
The comparison to Bethe networks with identical connectivities $c$ 
shows that both networks carry the same current 
$J(\rho)$ at very low and very high densities. However, they behave 
very differently at intermediate densities. We observe that 
(i) currents in Poissonian networks are significantly lower 
than in the corresponding Bethe networks,
(ii) even on the MF level, the current $J(\rho)$ no longer possesses particle-hole symmetry ($\rho \leftrightarrow 1\!-\!\rho$), and
(iii) the density at which the highest current is achieved lies below 
half-filling, and progressively reduces with connectivity $c$.
The most striking difference, however, is the absence of a 
plateau in $J(\rho)$ for the Poissonian network. 
In contrast to Bethe networks, this suggests that no SP segments 
involving domain walls arise over any extended density range.

\begin{figure}
 \begin{center}
   \includegraphics[width=.45\textwidth]{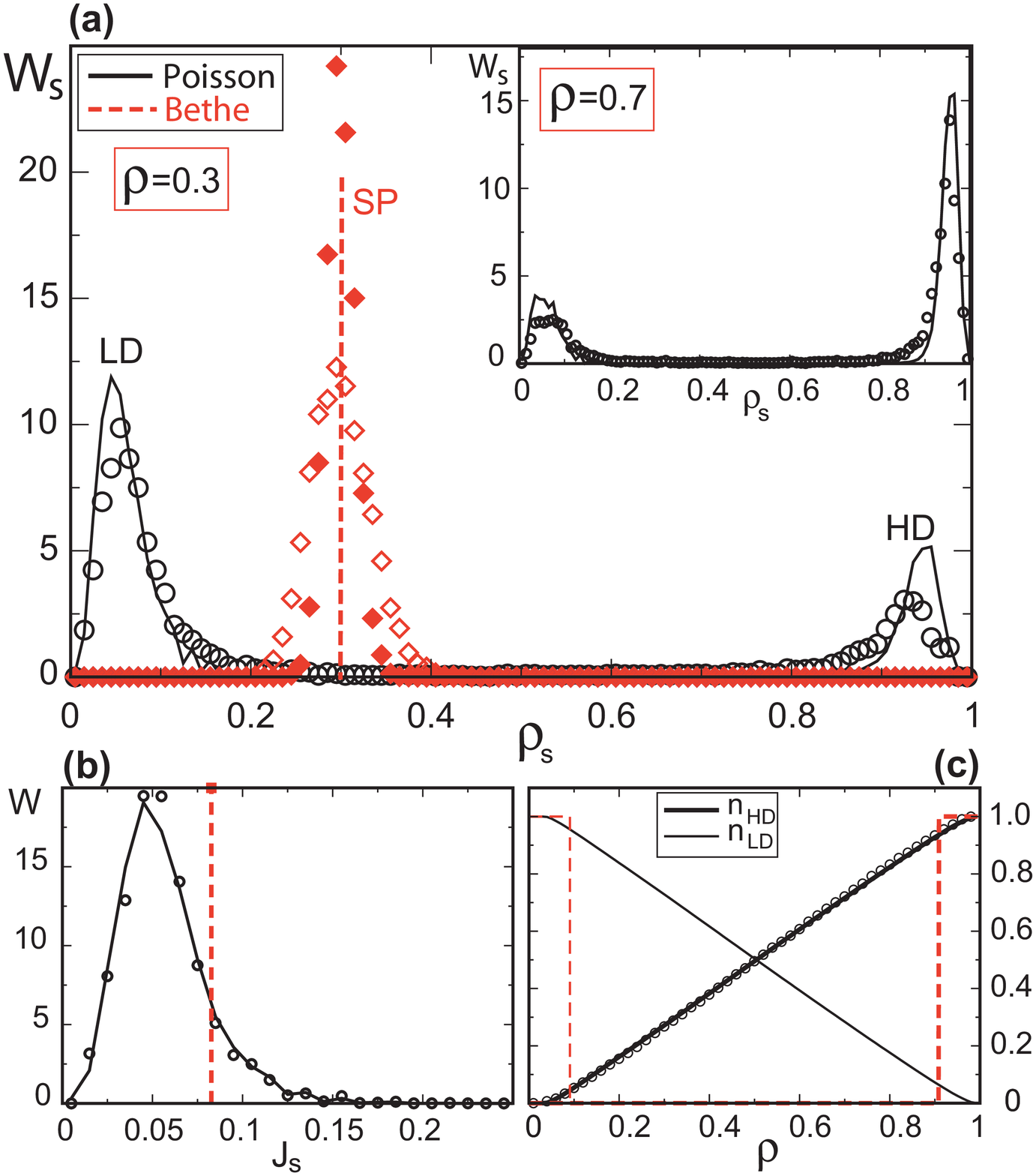}
 \caption{(color online). Distribution of segment properties for networks of 
   connectivity $c=10$, 
   comparing a Poisson network (black) to a Bethe network (dashed red).
   Symbols show simulation results, lines are mean-field predictions.
   (a) Distribution of segment densities $W_S(\rho_s)$ for an overall density 
   $\rho\!=\!0.3$ (inset: $\rho\!=\!0.7$). Convergence of the density peak to a delta peak is very slow in simulations due to collective fluctuations in the shock phase, closed markers correspond to fourfold longer runs than open ones.
   (b) Distribution of segment currents $W(J_s)$~for~$\rho\!=\!0.3$.
   (c) Fraction of segments in high ($n_{\rm HD}$) and low ($n_{\rm LD}$) density segments as a function of the overall density~$\rho$. 
 } 
 \label{fig:distributions}
 \end{center}
\vspace{-0.5cm}
 \end{figure}

A finer understanding is obtained by analysing how the transport features 
of individual segments are distributed across the network.
Consider first the distribution of segment densities, 
shown for a connectivity $c\!=\!10$ and an overall density 
$\rho=0.3$, see Fig.\ref{fig:distributions}(a). 
It is bimodal with peaks corresponding to segments 
carrying either a very low or a very high density. 
We are therefore dealing with two sub-networks of segments either 
in LD or HD phase, in contrast to the 
Bethe scenario where all segments are in SP at intermediate densities
(single peak in red). 
Interestingly, the distribution of the segment currents 
remains unimodal for Poisson networks, Fig. \ref{fig:distributions}(b),
thus putting a similar load on all segments.

The bimodal density distribution is also the key to understanding how the 
network adjusts to higher overall densities, by 
successively switching individual segments from a 
LD to a HD phase.
The inset in Fig.\ref{fig:distributions}(a), obtained for $\rho=0.7$, shows
that the typical densities of HD/LD segments are not significantly 
modified, whereas the proportion of the HD network grows 
at the expense of its LD counterpart 
as further particles are added.
This is further documented by the fraction $n_{\rm HD}$  ($n_{\rm LD}$) of edges in 
HD (LD) phases, Fig. \ref{fig:distributions}(c). For the Poissonian network 
$n_{\rm HD}$ is roughly equal to the overall density $\rho$,
thereby confirming that the change from LD to HD in irregular networks  
occurs progressively.  This linear behavior of $n_{\rm HD}$
is furthermore very robust with respect to variations of average 
connectivity $c$ (data not shown), implying that this picture remains 
valid for all connectivities $c$. 
\\

\paragraph{Transport on highly connected networks. --}
Significant insight into TASEP transport on {\it general} random networks
can be gained from analyzing the (MF) high-connectivity limit.
Ultimately all vertices constitute bottlenecks, taking their occupancy 
close to saturation. 
Therefore we expand the MF Eqs. (\ref{eq:MF}), posing  
$\rho_v = 1 - r_v/c$, with $r_v\sim \mathcal{O}(1)$.
The distribution of individual segment densities 
$\rho_s$  then becomes 
\cite{supplemental}
\begin{eqnarray}
  W_S(\rho_s) 
  =\sum_{k} \frac{p_{\rm deg}(k) \ k}{c}
  \int dr \, W_V(r) \,
  \delta\left(\rho_s-\rho_\pm\right)
  &&
\end{eqnarray}
where $p_{\rm deg}(k)k/c$ represents the degree
distribution of vertices when selecting a random segment, $W_V(r)$ is the
distribution of $r$, which follows from the MF Eqs. (\ref{eq:MF}).
The high (low) density value $\rho_+$ ($\rho_-$) are set by the
vertex saturation parameter $r/c$ relative to the
vertex degree $k$, as
\begin{eqnarray}
  \rho_\pm = 
  \left\{
  \begin{array}{lll}
    \rho_- = 1/k &\mbox{if}& r/c > 1/k \\
    \rho_+ = 1-r/c &\mbox{if}& r/c < 1/k \\
  \end{array}
  \right.
  &&
  \ .
\end{eqnarray}
This shows that the bimodal distribution $W_S$ in the segment densities,
with a fraction of segments in LD and a fraction in HD, is a general 
feature for complex irregular networks.  
This is particularly well illustrated in the strong connectivity 
limit $c\rightarrow \infty$, where $W_S(\rho_s)$ reduces to 
\begin{eqnarray}
W_S(\rho_s) = (1-\rho)\,\delta(\rho_s) + \rho\,\delta(1-\rho_s) \label{eq:dens1} \ .
\end{eqnarray}
The prefactors of the $\delta$ functions explain the linear behavior of $n_{HD}$,
and our above interpretation for Poissonian networks hence generalizes 
to general irregular random networks. 
\\
\paragraph{Conclusions and Outlook. --}
TASEP transport on closed random networks has been shown to lead to very different scenarios for regular and irregular topologies.
%
%
Despite the minimal character of TASEP, there may be direct implications:
the presence of bimodality, which we have shown to be robust, in
biological tracer experiments would make our findings directly useful 
for their interpretation.
%
But our results also raise interesting questions, such as 
the interplay of biological transport and crowding, and their possible regulation by 
the cytosqueletal network.

An important result of our study is that MF arguments lead to very good predictions for the global transport properties of closed networks, and provides a framework for their interpretation. 
Moreover, our numerical MF method gives access to system sizes currently beyond the reach of simulations. 
Generalizations of the approach to open random networks 
and other network topologies seem straightforward. 
In addition, we may also expect it to generalize to other transport processes, as long as the behavior of an individual segment is known and boundary-controlled, i.e. determined by the occupancy of the junctions it connects to.
\begin{acknowledgments}
IN thanks Francesco Turci for many fruitful discussions. We thank A.C. Callan-Jones for a critical reading of the manuscript and acknowledge support from ANR-09-BLAN-0395-02 and from 
 the University of Montpellier 2.
\end{acknowledgments}

\end{document}